\documentclass[conference]{IEEEtran}

\pdfoutput=1

\usepackage{color}
\usepackage{url}
\usepackage{mathtools}
\usepackage{amsmath} 
\usepackage{xspace} 
\usepackage{lineno}
\usepackage{tikz}
\usepackage[utf8]{inputenc}
\usepackage[ruled,linesnumbered]{algorithm2e}
\usepackage{listings}
\usepackage{tabularx}

\newcommand{\xbook}{\textsc{xBook}\xspace}
\newcommand{\ossobook}{\textsc{OssoBook}\xspace}
\newcommand{\archaeobook}{\textsc{ArchaeoBook}\xspace}
\newcommand{\anthrobook}{\textsc{AnthroBook}\xspace}
\newcommand{\excabook}{\textsc{ExcaBook}\xspace}
\newcommand{\palaeodepot}{\textsc{PalaeoDepot}\xspace}
\newcommand{\anthrodepot}{\textsc{AnthroDepot}\xspace}
\newcommand{\inbook}{\textsc{InBook}\xspace}

\newcommand{\ossoinst}{\textsc{OssoInst}\xspace}
\newcommand{\ossobookupdater}{\textsc{OssoBook Updater}\xspace}
\newcommand{\xbooklauncher}{\textsc{xBook Launcher}\xspace}

\newcommand{\obinit}{\textsc{OBinit}\xspace}

\definecolor{gray}{rgb}{0.4,0.4,0.4}
\definecolor{darkblue}{rgb}{0.0,0.0,0.6}
\definecolor{cyan}{rgb}{0.0,0.6,0.6}

\newcommand{\lstsetxml}{
  \lstset{
    numbers=left,
    numberstyle=\tiny,
    stepnumber=1,
    numbersep=5pt, 
    basicstyle=\footnotesize\ttfamily,
    columns=fullflexible,
    showstringspaces=false,
    commentstyle=\color{gray}\upshape,
    breaklines=true
  }
}

\lstdefinelanguage{XML}{
  morestring=[b]",
  morestring=[s]{>}{<},
  morecomment=[s]{<!--}{-->},
  stringstyle=\color{black},
  identifierstyle=\color{darkblue},
  keywordstyle=\color{cyan},
  morekeywords={xmlns,version,type,encoding,standalone} % list your attributes here
}

\makeatother

\begin{document}

\title{\vspace{-0.15cm}The Historic Development of the\\Zooarchaeological Database OssoBook and the \\xBook Framework for Scientific Databases}

\author{
\vspace{-0.45cm}
\IEEEauthorblockN{\textbf{Daniel Kaltenthaler and Johannes-Y. Lohrer}}\\
\IEEEauthorblockA{Ludwig-Maximilians-Universit\"at, Institut f\"ur Informatik,\\ 
Lehrstuhl f\"ur Datenbanksysteme und Data Mining,
Munich, Germany\\\vspace{4px}
Email: \{kaltenthaler,lohrer\}@dbs.ifi.lmu.de}
}

\maketitle

\noindent \textbf{\emph{Abstract}} --- \textbf{In this technical report, we describe the historic development of the zooarchaeological database \ossobook and the resulting framework \xbook, a generic infrastructure for distributed, relational data management that is mainly designed for the needs of scientific data. We describe the concepts of the architecture and its most important features. We especially point out the Server--Client architecture, the synchronization process, the Launcher application, and the structure and features of the application.}

~

\section{Introduction}

In general, software and its possibilities are developing to an ever more advanced level. The implementations are changing over time and new technologies must be considered and integrated. Different ideas and concepts of developers, and different expectations of customers must be taken into account when developing the application. Even though these approaches may differ from expected realizations, especially in the range of data gathering the requirements are quite similar in several scientific areas.

It is often the case that different disciplines develop their own software solutions to gather and manage own data specifically for their own need. However, this causes the decisive disadvantage that new features, which can be applied to several disciplines, have to be implemented multiple times for each of the individual solutions. This is not only a huge temporal, but also a financial expenditure.

This situation became especially apparent during the development of \ossobook \cite{ossobook}, a database for zooarchaeological findings. Other disciplines in the archaeological context also gather data with similar methods. Of course, these differ content-related, but the requirements on the basic features strongly overlaps with the functionalities of \ossobook. This is not limited to the archaeological context, other scientific disciplines could also benefit from similar solutions.

However, in other archaeo-related disciplines (like archaeology, anthropology, archaeobotanic, etc.) there is also a necessity of gathering data which consists of similar workflows like in the zooarchaeology. In consequence, the idea of the \xbook framework developed out of this context. In the \xbook framework features are developed centrally and are provided to all applications that are incarnations of the framework. New features do not have to be implemented individually, however custom extensions are still possible. In case of errors and bugs it is not necessary to fix them in each application, they can be fixed centrally. Thereby, it causes the creation of similar structures in the gathering and analysis of data. As of now, the \xbook framework enabled the development and usage of a number of archaeo-related applications, like \ossobook, \archaeobook \cite{archaeobook}, \anthrobook \cite{anthrobook}, \excabook \cite{excabook}, \palaeodepot \cite{palaeodepot}, \anthrodepot \cite{anthrodepot}, and \inbook \cite{inbook} (cf. Section~\ref{applications}). At the same time, the \xbook framework is not limited to the requirements of the archaeo-related context, it can be applied to many other scientific application as well.

In this technical report we describe the development process of the \ossobook application and the \xbook framework in the recent decades. We first describe the origins of \ossobook in Section~\ref{origin} and explain the further development of the application and the extraction of the \xbook framework in Section~\ref{furtherdevelopment}. Finally we show the basic, most important features of \xbook in Section~\ref{features}.
\section{Origin of \ossobook} \label{origin}

Below, we describe the original development of the \ossobook application since 1990.

\subsection{First \ossobook version in dBASE}

\begin{figure}[h]
\centering
\includegraphics[width=\columnwidth]{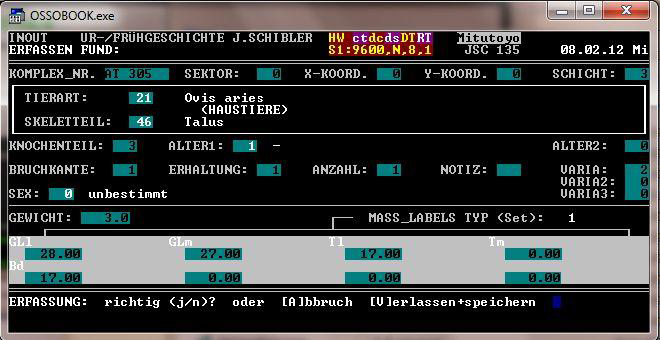}
\caption{The input mask in \ossobook 1.0. \cite{kaltenthaler2012} \cite{lohrer2012} \label{fig:dbase_ossobook1}}
\end{figure}

The first version of \ossobook was originally released in 1990 by J{\"{o}}rg Schibler and Dieter Kubli of the University of Basel, Switzerland. The technical basis was dBASE\footnote{dBASE's underlying file format, the .dbf file, is widely used in applications needing a simple format to store structured data.}, a file based database application for computer systems running the operating system DOS. The exclusively in German language published \ossobook database enabled the recording of zooarchaeological data. Five input fields for archaeological information, eleven input fields for zooarchaeological data, and for each skeleton element eight further input fields for the recording of measurements were provided (cf. Fig.~\ref{fig:dbase_ossobook1}).

%%%%%%%
%%%%%%%

\begin{figure}
\centering
\includegraphics[width=\columnwidth]{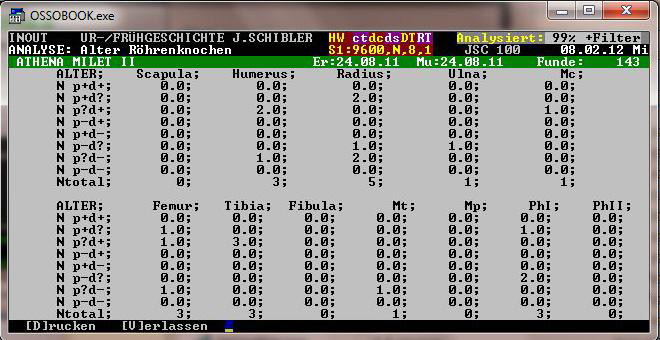}
\caption{The analysis of the age of long bones in \ossobook 1.0. \cite{kaltenthaler2012} \cite{lohrer2012} \label{fig:dbase_ossobook2}}
\end{figure}

Besides the data input, the early version of the application already provided the possibility to implement simple data analyses. Three analyses were available for skeleton element representations: One for the age of long bones (cf. Fig.~\ref{fig:dbase_ossobook2}), one set of analyses for bone parts, and one last analyses for measurements of bones. The results of these analyses could be saved to extern files which could be viewed and edited with spreadsheet like Microsoft Excel or Open Office Calc (today: LibreOffice Calc).

%%%%%%%
%%%%%%%

Besides to the database itself, this version also provided an editor called \ossoinst which allowed defining custom numerical codes for different data inputs, e.g. the mapping of a species ID to a species name. This offered the advantage, that each database could be used with individual data, each user could easily extend the list of available species.

The archaeological data and the corresponding mappings were saved as ASCII files on the local hard disc drive of the computer, or on floppy disks. This made the sharing of data complicated and difficult, and therefore was rarely practiced. Nethertheless, \ossobook already allowed the translation of specific texts of input values to other languages by using \ossoinst. First partial translations (especially to English and French) were already possible and done. \cite{schibler1998}

\subsection{Conversion to Java} \label{conversion_to_java}

At the latest with the release of Microsoft Windows XP in 2001, the operation system DOS fades into the background. The technique of the dBASE database became increasingly obsolete. Even though the application is still running on modern operation systems (e.g. in the command-line interface on Windows computers, or the system console in other operation systems), the usability and optical presentation of \ossobook was no longer up-to-date. The disadvantage that data could only be saved on the local computers, also contributes that a new, enhanced version of \ossobook should be developed.

The new version of \ossobook was initialized by Christiaan H. van der Meijden of the Tierärztliche Fakultät\footnote{\url{http://www.vetmed.uni-muenchen.de}}, together with the Institut für Informatik, Lehrstuhl für Datenbanksysteme and Data Mining\footnote{\url{http://www.dbs.ifi.lmu.de}} of the Ludwig-Maximilian-Universität München, Germany. The application was converted to the object-oriented and platform-independent programming language Java. On the servers of the university, there was installed a single, global MySQL database which should be used by the employees of the institute. They used the client to connect to the server and directly work with the data of \ossobook on the global database. \cite{lamprecht2008}

Besides, the mapping of IDs to values for specific fields was adopted, but the functionality to add or change them manually was removed. The users worked with standardized, predefined values for the necessary input fields. This should improve the comparability of the entries saved in the database. These mappings are now called ``Code Tables'' in the application.

\begin{figure}[t]
\centering
\includegraphics[width=\columnwidth]{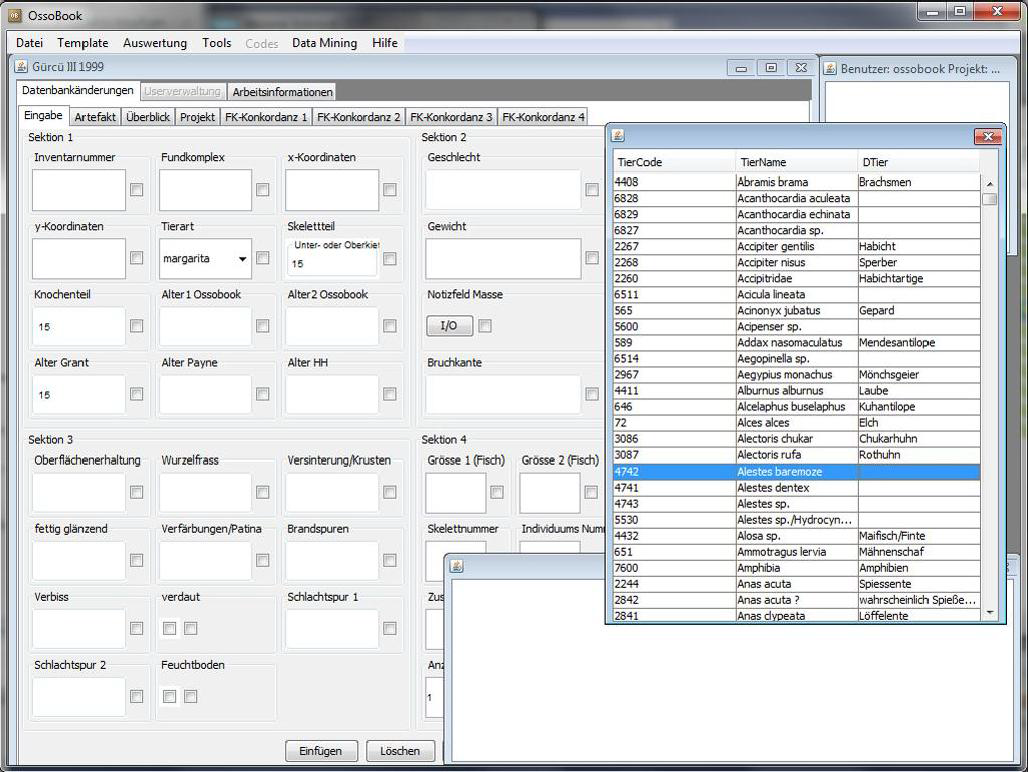}
\caption{The input mask in \ossobook 3.4. \cite{kaltenthaler2012} \cite{lohrer2012} \label{fig:ossobook_3-4_inputmask}}
\end{figure}

In combination with the port of the application to Java, \ossobook got a graphical user interface for the first time. As shown in Fig.~\ref{fig:ossobook_3-4_inputmask}, the input fields were arranged in four sections and offered first input assistances, e.g. by using selection boxes for predefined values. Statistical information about the data sets and simple analyses were displayed in several tabs, which also provided more space for further input possibilities. 

Version 3.4 was the first Java version of \ossobook and was released in 2007.

\subsection{First implementation of a synchronization}
Until this date the users of \ossobook could only connect and work directly on the global database on the servers. Originally this was only possible with connections within the network of the university, later a tunnel enabled the connection from other places as well.

In 2008, the first implementation of a synchronization in \ossobook allowed the entry of data in a local database of the clients. The users could enter their data remotely without any connection to the network and later synchronize it to the global database. The implementation and the full development of the synchronization is described in detail in Section~\ref{xbook_synchronization}.

%%%%%%%
%%%%%%%

\subsection{Redevelopment of the application}

At this time, the authors of this report were tasked with the further development of \ossobook. However, this Java version of \ossobook caused big problems for the development and usage. One could see that the application was only a port and did not use the potential of the object-orientated programming language Java. Most of the input fields were not very intuitive due to having to work with numerical codes in general, which meaning had to be looked-up in external spreadsheet or PDF files. Besides, the application included a lot of errors and bugs which were not displayed in the graphical user interface of the application. In some cases these problems made the data input impossible and made the application crash.

From the point of view of the authors it was nearly impossible to implement the requests of the zooarchaeologists and to add new features. The first tries to integrate new elements into the application made already clear that it is more reasonable to re-implement the application from scratch instead of trying to continue developing for the current version. The main problem of the further development of the current version was the very static program code elements that did not allow adding new input elements to the input mask. This static design also made a object-orientated design using inheritance impossible. Also the programming code was only sparsly commented and Javadoc comments were missing for the most parts. In addition, the names of the methods were not intuitive, so new developers would need a long familiarization time to be able to develop the application.

It was decided that the database scheme of \ossobook and the \ossobook client should be newly developed considering the Model--View--Controller architecture \cite{kaltenthaler2011} \cite{lohrer2011}. We put a lot of emphasis for future features being able to be fast and dynamically integrated to the application to enable a simple and resource-efficient further development. To avoid data loss and to be able to continue using the data of the previous \ossobook version, we wrote a script that converts the old database scheme into the new one.

\begin{figure}[t]
\centering
\includegraphics[width=\columnwidth]{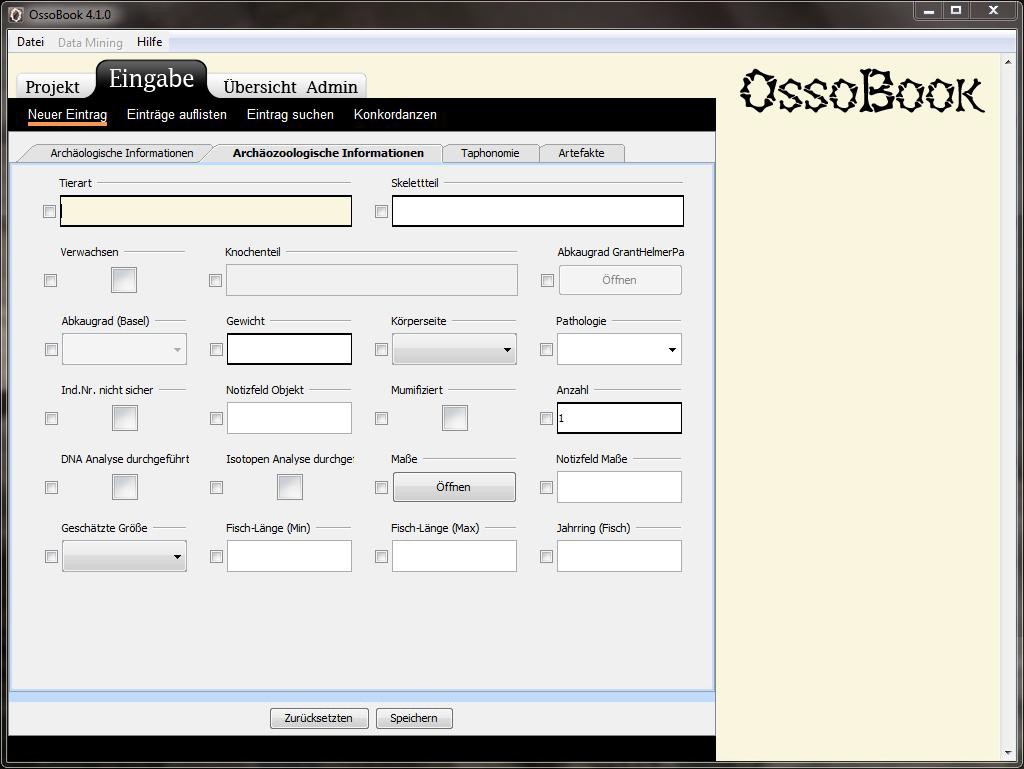}
\caption{The input mask in \ossobook 4.1. \cite{kaltenthaler2012} \cite{lohrer2012} \label{fig:ossobook_4-1_inputmask}}
\end{figure}

In autumn 2011, the version 4.1 of \ossobook was released that included the same features than the previous version of the database application at first, but was more flexible in the usage for the users and developers. A screenshot of this version can be seen in Fig.~\ref{fig:ossobook_4-1_inputmask}.

%%%%%%%
%%%%%%%

\section{From the Application OssoBook to the xBook Framework} \label{furtherdevelopment}

From this point of time, \ossobook was ready for the usage of the scientists. At the same time it is assured that further development and future adjustments to the database are possible.

Since the gathering of data is an essential task of the work archaeo-related disciplines, other disciplines got also interested in the architecture of \ossobook. While the workflow process is similar in all disciplines (archaeology, anthropology, zooarchaeology, palaeobotanic, palaeontology, etc.), the collected data is different in each special field. An individual database solution based on the \ossobook architecture would greatly support the scientists in their work.

So we set the challenge to provide a generic solution for supporting the scientists in all disciplines, that is as customizable as possible to allow all required information about the specific data to be gathered. Therefore, we used the basic architecture and features of \ossobook and extracted \xbook, a generic framework including the common and basic features for a database for archaeo-related disciplines.

In this Section, we describe the most important functionalities that are features of the \xbook framework.

%%%%%%%%%%%%%%%%%%%%%%%%%%%%%%%%%%%%%%%%%%%%%%%%%%%%%

\subsection{Input fields and input mask}

The input fields were strongly enhanced and extended. Previously, there had been only four basic types of fields available: Text, numeric values, check boxes, and Code Tables. Several new types of fields were integrated which can be reused for new input fields, e.g. combo boxes for values and IDs, multi selection data, buttons to open panels for more complex data inputs, date and time choosers, etc. Furthermore, several individual input elements were added that were specifically implemented for single data elements of \ossobook. Especially the input fields for species, skeleton elements, measurements, wear stages, and the bone elements benefited from the individual input possibilities.

\begin{figure}[t]
\centering
\includegraphics[width=\columnwidth]{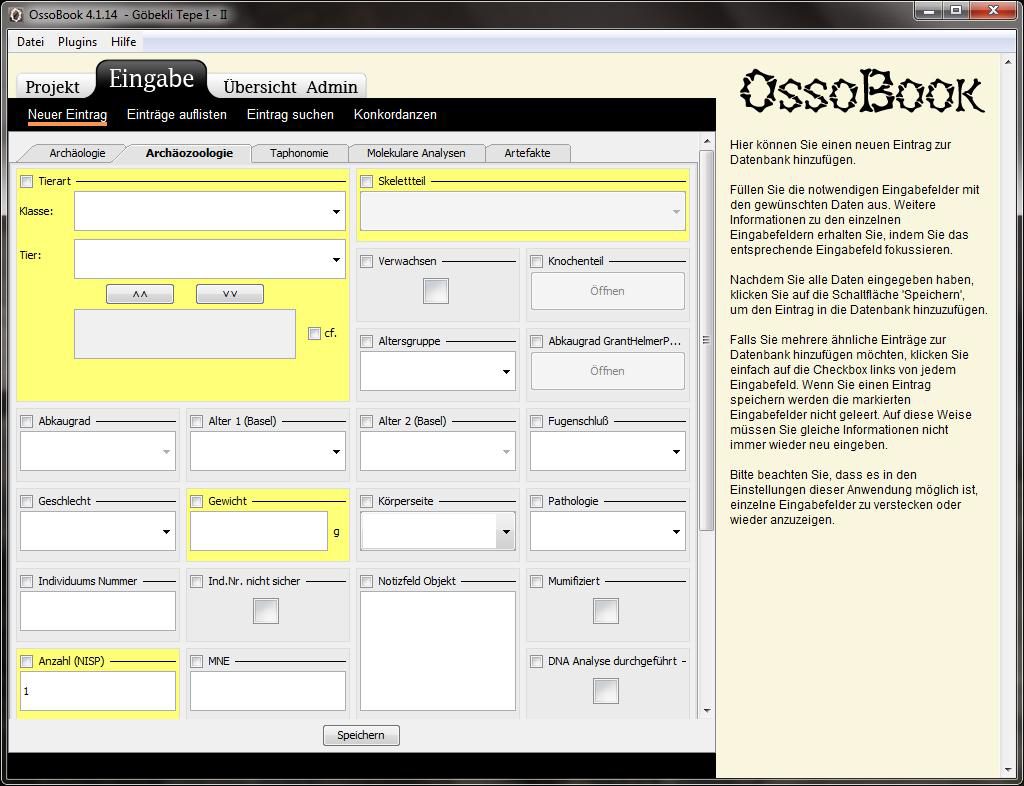}
\caption{The input mask in \ossobook 4.1.14. \cite{kaltenthaler2012} \cite{lohrer2012} \label{fig:ossobook_4-1-14_inputmask}}
\end{figure}

Also the visual presentation of the input mask was updated, as shown in Fig.~\ref{fig:ossobook_4-1-14_inputmask}. Besides the arrangement of single elements inside an input element, they were wrapped with a visible box that was able to be colorized dependent on different states. A mandatory field, that has to be filled before saving the entry, is highlighted with a yellow background color. If an input is not valid in a field, this is signaled with a red background color. Further enhancements in the graphical user interface were also added, e.g. a box for temporarily displaying text like warnings and errors as a feedback for the user.

%%%%%%%
%%%%%%%

%%%%%%%%%%%%%%%%%%%%%%%%%%%%%%%%%%%%%%%%%%%%%%%%%%%%%

\subsection{Update procedure}

To enable a dynamic development of the application and the version-independent use of the synchronization it was necessary to keep the database and the program version up-to-date. Only if the local and global database match the same database scheme, the synchronization is able to exchange data correctly. So an update procedure was integrated, that consisted of three steps:

When the user logs in, the program version of the local client is checked if it is up-to-date. If not, the \ossobookupdater, a small helper application, was 
% automatically 
started which let the user update the program files by executing the update process.

Then the update process updated the database scheme and the data itself to the current version, if necessary. In the program code it was defined which database version is required for the current version. If the current, local database version did not match with the database version in the program code, the necessary SQL queries are loaded from the server and executed. This was done recursively until the database versions matched.

Finally the Code Tables were updated. To guarantee a consistent data structure, it was also important that the mapping of the values to the corresponding IDs is the same on each client and on the server. So the synchronization was extended with a method that updated the Code Tables in the local clients.

\subsection{Plug-in interface}

At that time, \ossobook provided an interface for plug-ins which was used for the integration of different sets of analysis that were developed by students of the Ludwig-Maximilians-Universität München:

\begin{itemize}

\item A module for the analysis of age distribution \cite{kaltenthaler2012}
\item a module for the cluster analysis of measurements \cite{lohrer2012}
\item a plug-in for similarity search on multi-instance objects \cite{danti2010}
\item a plug-in for the execution of sample data mining methods \cite{tsukanava2010}, and
\item a module offering some analysis methods for zooarchaeological data \cite{neumayer2012}.

\end{itemize}

These plug-ins were able to be run directly in the application and used with the data from the database. Each plug-in could be imported to the application by simple copying the corresponding jar-file into the plug-in folder of \ossobook.

%%%%%%%%%%%%%%%%%%%%%%%%%%%%%%%%%%%%%%%%%%%%%%%%%%%%%

\subsection{Database Identification} \label{dbid}

For the synchronization and the possibility to work offline, it was essential to differentiate data sets that were entered on different computers. The problem is that one single ID for the data sets is not sufficient because the same ID could be assigned on different computers several times which would cause errors in the synchronization process. This was solved by the addition of a new column called Database ID for each entry. 

Storing and handling of the Database ID required several iterations to prohibit errors and problems with different aspects of user interaction:

The first iteration considered a file called \obinit which was a short SQL script that was generated during the registration process and sent to the user. The file included the username and password and additionally assigned a unique Database ID to the local database. The main issue of this solution was that the \obinit file was necessary to initialize the database, but if the users installed the application on two different computers and used the same \obinit file, the identical Database ID was used for both computers. Furthermore, the users had to save the email and the \obinit file because it was not possible to recover the data once the information is lost. This solution also bound the password to the \obinit file, which was also a reason for why the password was not editable.

The second iteration approached these main issues. The assignment of the Database ID was realized by the server while initializing the database. So every time a new database was installed and initialized on a different computer, the server was queried for a new Database ID which was used for the local database. In theory, this approach solved the problems with different computers, however some users created local backups of the application folder which also includes the local database. So it occurred again that the combination of identical entry IDs and database numbers were used when the users restored the application from the backup.

The final iteration closes this gap by defining that the application could not be installed on any folder anymore and additionally saving the database ID in the registry. Now the \xbooklauncher (cf. Section~\ref{launcher}) installs the application data and the database in a folder which grants the logged-in user reading/writing access, e.g. in Windows we use the AppData folder. When the application is started, the Database ID inside the database is checked against the ID saved in the registry. If they do not match, or is not yet available, a new Database ID is issued, which is then saved again in both the database and the registry.

%%%%%%%%%%%%%%%%%%%%%%%%%%%%%%%%%%%%%%%%%%%%%%%%%%%%%

\subsection{Registration}

Originally, the users could register for an \ossobook account at a password protected homepage only. The users had to enter an email address and got an email including the username, password, and the necessary \obinit file (cf. Section~\ref{dbid}). This information was sufficient to work with the application, however, common mechanics like editing the user name or email address, or change/recover the password were not supported.

Later we changed this system to a more modern approach. The registration was moved directly into the application. At the login screen we added a button to register, where the users can enter some basic information: User name, first name, last name, email address, and a password. Now, there is no user restriction anymore, everyone can register and use the application. Once registered, the users can login to the application without the need of a \obinit file -- due to the reasons described in Section~\ref{dbid}. Furthermore, \ossobook was extended with profile settings where the users can manage their provided data. Especially the application was extended with a feature allowing the users to change and to recover their passwords.

%%%%%%%%%%%%%%%%%%%%%%%%%%%%%%%%%%%%%%%%%%%%%%%%%%%%%

\subsection{Server--Client architecture}

Having a reliable Server--Client infrastructure is an important requirement to be able to synchronize and backup data. This also helps to ensure no unauthorized changes are done, e.g. by a hacked client. In our case the Server--Client architecture has to handle different scenarios. The first one is the registration and login process. For this it is necessary to connect with the server from anywhere. After the user logged in, the server has to check if the client is up-to-date or first has to be updated. For this the database scheme has to be sent to the client along with values of the code tables. After the version check is completed, the main task of the server is to handle the synchronization requests from the client. These use cases require the communication to handle a variety of dynamic and versatile data. Additionally -- since client and server are implemented in different programming languages -- built-in serialization tools like the Java Serialization \cite{oracle2017} can not be used. In an environment where multiple users can create, edit, and share their data, it is important to have a managed architecture that can be accessed and used from everywhere without any restrictions.

\subsubsection{Challenges}

A Server--Client architecture faces many challenges. Many of these are common in every Server--Client application, but some are very specific to the needs of the \xbook framework.

\begin{itemize}

\item \textbf{Security:}\\
Prevent unauthorized access. A user must only be able to access the data he has the rights to. Unauthorized access must be prevented.

\item \textbf{Availability:}\\
The server must be available from everywhere. Using a Socket-based architecture \cite{sockets2018} generally requires the usage of ports, which have to be manually opened by an administrator in a firewall-protected secure environment. However, the opening of a port is not possible in every working environment because of strict regulations which forbids users to communicate with servers on other ports than 80 (HTTP) or 443 (HTTPS), for example in offices of state authorities or some institutes. Therefore, we had to find a solution how to make a connection from the client to the server possible in spite of restrictive firewall policies and how to use the available ports for the \xbook server to accept requests.

\item \textbf{Scalability:}\\
The server must work for single and also multiple users at the same time. This is true for most Server--Client architectures, still multiple users working on the same server must be able to work simultaneously and not having to wait for one request to be completed, until the next one is carried out.

\item \textbf{Flexibility:}\\
The server should run independent of the Book without any knowledge of data scheme inside the server. Therefore the specifics of each individual Book must not be hard coded inside the server application, but dynamically loaded from a configuration file or the database.

\end{itemize}

\subsubsection{Evolution} 

The first synchronization of the data in \ossobook was handled by directly connecting to the database on the server from the client application. This connection required a manually entered passphrase, which was given out with the registration, but was identical for all users. Additionally -- apart from the client itself -- no further checks for authorization were made. To address these issues, a C++ server application was created with the development of the \xbook framework, which now was the communication partner of the client application. The server is connected to the database and analyzes incoming requests, if the user has the authorization. If this is the case, the server carries out the command and sends a confirmation back to the client together with data, which was retrieved by this request. To allow multiple users to work simultaneously, both a Thread Pool \cite{garg2002} \cite{goetz2002} and a Connection Pool \cite{goll2014} were used. The communication between server and client was handled via sockets with a custom serialization of all objects that were transmitted. While this architecture provided a fast, secure, scalable, and flexible way to communicate with the server, it became clear that -- due to the nature of sockets -- it could not be guaranteed that the connection can be established behind proxy servers that only allow certain ports. Therefore the communication had to be moved to a different type of protocol.

To solve the problem with proxies restricting certain ports, the communication had to be done over ports which are not restricted by most proxies. These are usually port 80 (HTTP) and port 433 (HTTPS). Of course, the possibility remains that certain IP addresses are blocked. However, it would really get into hacking to get around this, we did not explore this possibility further. The server which is running the server application is also running an Apache server. This is used to distribute the \xbooklauncher and to download the files required to start the individual Books. Besides it hosts the \xbook Wiki\footnote{\url{http://xbook.vetmed.uni-muenchen.de/wiki/}}, a MediaWiki that provides helpful information. So it was not possible to change the port to 80 or 433 the old C++ application was listening on. A new server application was required that does not conflict with the Apache server, but can run alongside it. Many different web applications would allow this, but PHP was chosen as a scripting language, since it does not require additional configuration of the server.

\subsubsection{Communication}

In a traditional web service, the user would enter an URL in their browsers. The web server would then analyze the request and return a website with the requested information. Since in our case, the client has to communicate directly with the server, the response must not be human readable, but interpretable by the client. Because the client was already able to communicate with the old C++ server, the serialization on the client side was already available and working. Still, it had to be modified to be able to communicate with a PHP server. On the other hand, PHP is not designed to work with serialized objects, but to load a script with some parameters, and then return and display results. 

There are several possibilities to realize a cross-platform serialization. One is JSON \cite{json1} \cite{json2}, a lightweight, text-based, language-independent data interchange format. Since JSON uses a human-readable format it has the disadvantage of data overhead \cite{jsoncompression}. There are ways to optimize the transmission size, e.g. XFJSON transforms the JSON format into a binary-hex form that is additionally encoded and decoded \cite{xfjson}. Considering that there is no necessity to be able to read the transmitted data and that we expect a huge amount of data sets for single projects, we focus on bandwidth-efficient solutions. So we need an alternative that is not human-readable. However, a solution like FlatBuffers \cite{flatbuffers} was not published at the time of the implementation of our serialization method. Protocol Buffers \cite{protocolbuffers} do not support PHP at all. BSON in direct comparison with Protocol Buffers can give an advantage in flexibility but also a slight disadvantage in space efficiency due to an overhead for field names within the serialized data \cite{bson}. However, BSON is mainly used as a data storage and network transfer format in MongoDB\footnote{\url{https://www.mongodb.com}} databases. Therefore, we would have had to distribute all libraries for MongoDB which seemed unreasonable, since there is no stand-alone implementation. As no good alternative for our serialization was available at the time of implementation, we had to implement an own solution.

Since the communication with a PHP server is asymmetrical, as requests and results are not communicated the same way, it was necessary to split the serialization in two parts. The first part consists of the data transmitted to the server. For this the request is serialized to a string which is then appended to the requested URL with the HTTP POST method. This allows the PHP server to read the data and deserialize the string back to the request, which is then carried out. After the server completed the request, the result is serialized again and the resulting string is displayed as the content. This is read by the Java client and is returned deserialized.

The communication is done with a message object. The message object holds the type of the request, e.g. synchronization, login, register, and additionally a list of further data. All classes that can be added as data are instance of the interface \verb|Serializable| which has methods to serialize and deserialize itself. For each request the type and amount of parameters of the data that is sent is predetermined. Of course, the data itself is not known beforehand.

The serialization requires special classes that implement the \verb|Serializable| interface even for basic data types like \verb|String| or \verb|Integer|. Currently 16 different classes are used that can serialized. These are mostly required for the synchronization and the initialization of the database scheme.
To secure the communication HTTPS is used.
%{\color{red}This required little changes,.. (really interesting?)}
To ensure independence of the database model of the individual Book, the server should hold no information about the specific Book, apart from information of the corresponding database to use. Information about the tables to include in the synchronization are saved in tables inside the database. This allows these tables to be dynamically adjusted, without the need to update the server. If a request is carried out, these tables are checked if and which columns can be accessed.

%%%%%%%%%%%%%%%%%%%%%%%%%%%%%%%%%%%%%%%%%%%%%%%%%%%%%

\subsection{Launcher} \label{launcher}

A very important part of the \xbook databases is the \xbooklauncher. Over the years, the application took on an increasingly important role. It developed from a simple updater application, that was executed if a new \ossobook update was available, to the central place where all Books can be installed, updated, and started independently. 

\subsubsection{\ossobookupdater}

\begin{figure}[t]
\centering
\includegraphics[width=1\columnwidth]{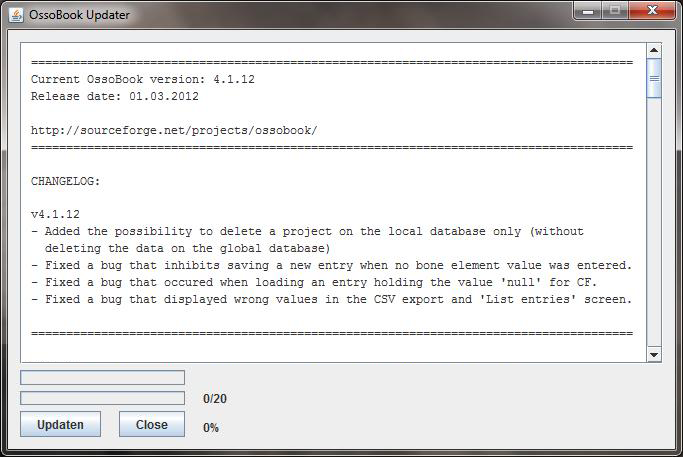}
\caption{The \ossobookupdater allowed the user to update the \ossobook application. \cite{kaltenthaler2012} \cite{lohrer2012} \label{fig:ossobook_updater}}
\end{figure}

The first idea of the \ossobookupdater was born through the necessity to allow the user to update the program. This was required because the latest version of \ossobook was required to be able to work in online mode and to communicate to the server, because the local database scheme must be identical with the global scheme on the server when synchronizing data. So the update process is a frequent process that has to be executed with each minor update of the application. A screenshot of the \ossobookupdater can be viewed in Fig.~\ref{fig:ossobook_updater}.

%%%%%%%
%%%%%%%

We wanted to avoid that the users had to run and install the update process manually. They should not have to go to the website, download the latest version, and install the files in the appropriate directory. The \ossobookupdater was automatically called when it was detected that the local program version was out-of-date when the user tried to connect to the global database. The \ossobookupdater had a list of files that had to be checked for updates and compared them to the files available at a specific website. Since \ossobook could be installed in an arbitrary directory, the \ossobookupdater had to use this directory to update the files. For this the updater was also in this directory and was updated by \ossobook before the updater was executed.

To prevent different instances being run on one single computer, we had to specify the directory in which \ossobook is located. This guarantees -- together with the ``DatabaseID'' (see Section~\ref{dbid}) -- that the instance on a single computer is both unique and can be identified uniquely. To avoid permission problems, a directory had to be used where writing permissions are guaranteed for the users. For Windows environments we chose the ``AppData'' directory. This allowed users to easily install \ossobook on one computer, synchronize their data, and continue to work on a different computer.

Instead of updating the files in the directory of the updater, the updater became a independent file, that from now on was called \xbooklauncher, since it also served as the entry point for the application. So instead of directly starting \ossobook, now the users had to run the \xbooklauncher which then checked if all files are up to date and then allowed the execution of \ossobook. 

\subsubsection{Development of the \xbooklauncher}

\begin{figure}[t]
\centering
\includegraphics[width=1\columnwidth]{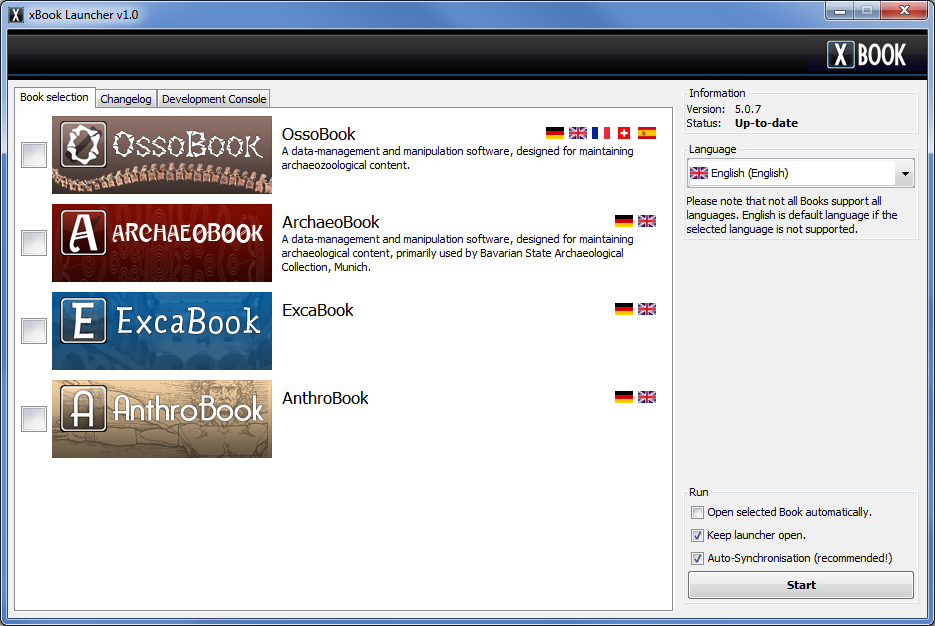}
\caption{The first \xbooklauncher 1.0 with the selection of four different Books based on the \xbook framework.
\label{fig:launcher_1-0}}
\end{figure}

With the development of more and more Books for different areas of work, the requirements for the \xbooklauncher changed and -- to avoid the need of several launcher applications for several databases -- had to be adjusted to support more than one database. Therefore, the \xbooklauncher was extended with a Book selection. Each supported Book was represented as an own row in the selection displayed by an application icon, the application name, a short description of the database and the supported languages. The user could select the desired database and execute it. Furthermore, the \xbooklauncher was extended with general settings that affects all Books (like the language selection and the selection of the automatical or manual synchronization) and a frame to output the messages of the development console. The update functionality was still available, which updates all Books at the same time. A screenshot of the version 1.0 of the \xbooklauncher can be viewed in Fig.~\ref{fig:launcher_1-0}

%%%%%%%
%%%%%%%

However, it became difficult to use one single launcher for all available Books. Some of the Books are scientific databases which are publicly available, others are local databases for the inventory of findings in museums or state collections. So not all Books should be accessible in public, also the users do not need to have listed all existing Books in their launcher. The previous solution needed to create an own instance of the \xbooklauncher for almost every Book, a roundabout way for the increasing number of Books. Furthermore, it became necessary that the different databases do not need to be hosted on the same server any longer, the single Books should also be supported to save their data on their own servers. So the structure of the \xbooklauncher was renewed again.

Therefore, the \xbooklauncher was extended to enable adding single Books dynamically to the list of Books. The users can enter a valid URL where the configuration file of the corresponding Book is saved. The configuration file is defined to be named ``Book.xml'' which holds all necessary information for the corresponding Book. This includes especially information that are used in the launcher to display the information about the Book (like application name, application id, multi language descriptions, etc.), but also defines the file that should be executed when running the Book and the location of the data that is required when installing or updating the application (cf. Algorithm~\ref{alg:bookxml}).

%%%%%%%%%%%%
\lstsetxml
\begin{algorithm}
\begin{lstlisting}[language=XML]
<?xml version="1.0" encoding="UTF-8"
        standalone="yes"?>
<book>
  <displayName>OssoBook</displayName>
  <id>ossobook</id>
  <fileName>OssoBook.jar</fileName>
  <author></author>
  <bookColor>#624D47</bookColor>
  <sidebarBackgroundColor>#D5CECA</sidebarBackgroundColor>
  <defaultLanguage>en</defaultLanguage>
  <downloadLocation>
    http://xbook.vetmed.uni-muenchen.de/books/ossobook
  </downloadLocation>
  <languageFiles>
    <entry>
      <key>de</key>
      <value>
        <entry key="description" value="[...]" />
        <entry key="descriptionShort"
                 value="[...]" />
        <entry key="author" value="[...] "/>
      </value>
    </entry>
    <entry>
      [...]
    </entry>
  </languageFiles>
  <languages>en</languages>
  <languages>de</languages>
  <languages>fr</languages>
  <languages>es</languages>
  <languages>de_ch</languages>
</book>
\end{lstlisting}
\caption{The (shortened) structure of the \emph{book.xml} file that shows the configuration for \ossobook.}
\label{alg:bookxml}
\end{algorithm}
%%%%%%%%%%%%

This way the \xbooklauncher can be used as a central platform to manage all available Books, independent on their location. Especially the users benefit from this architecture, they can configure their launcher as they wish and do not have to install/integrate Books they do not work with at all. Additional, the single Book applications can now be developed and hosted completely separately. Earlier all Books had to use the same program version because they all depended on the same update files. Now every Book can be developed independently from the other existing Books and must not necessarily be updated to the latest version of \xbook. This got more important with more and more different Books being published by different institutions and developers.

\begin{figure}[t]
\centering
\includegraphics[width=1\columnwidth]{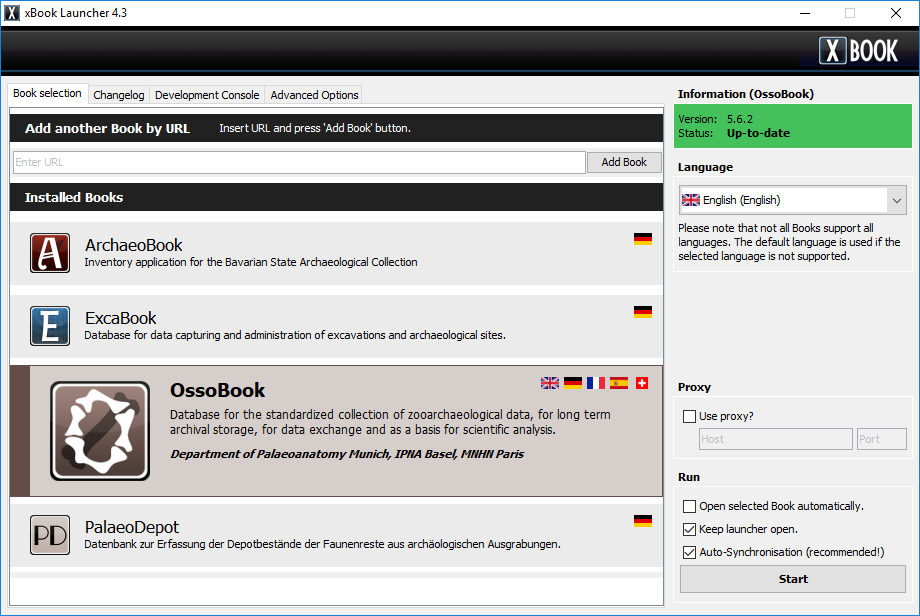}\vspace{-4pt}
\caption{The further developed version \xbooklauncher 4.3. \label{fig:launcher_4-3}}
\end{figure}

In addition, the \xbooklauncher was extended with two technical features. With the increasing amount of users, more users required the usage of a proxy server to be able to connect to the server. For this the \xbooklauncher was extended with a possibility to enter proxy information. Furthermore it becomes apparent that a lot of users work on computers with a limited amount of computer memory which was not sufficient for using some of the features of \xbook, like exporting the partial huge amounts of data. To solve the problem, the \xbooklauncher was extended with advanced options where the users can assign more RAM to Java applications, and therefore for the Book. A screenshot of the current \xbooklauncher version 4.3 can be seen in Fig.~\ref{fig:launcher_4-3}.

%%%%%%%
%%%%%%%

%\subsection{Database Connection?}
%{\color{red}-intro
%Datenverbindung notwendig 
%Datenbankschema muss bekannt sein.
%Viele spalten gleichen typs. 
%REusability von funktionen auf einzelne spalten -> ColumnTypes
%Unterschiedliche ARten von TAbellen. Projekt, Inputunit, und untertabellen. Selbe funktionen aber anders.
%Ziteate: Design Patterns
%verbindung über manager. 
%}
\section{Features of the \xbook framework} \label{features}

There are many different areas in the archaeological field of science which struggles with the same problems concerning the collection of data. \ossobook is a database for zooarchaeological findings, but similar databases were also demanded for other areas, like anthropology and archaeology. Instead of individually implementing a new database for each area, we decided to extract the features from \ossobook into a new framework called \xbook.

This framework should provide all basic functionalities of the application and provide them to each incarnation, which are called ``Books''. 

\subsection{Synchronization} \label{xbook_synchronization}

Since the conversion of \ossobook to Java, the collaboration in gathering and maintaining zooarchaeological data is an important part of the concept. Initially the data of the application was saved in one global database on the server of the university (cf. Section~\ref{conversion_to_java}). The employees of the collaborating institutes could connect and work on this database. There, they could enter, view, and edit the zooarchaeological data directly on the server. This made it possible to work together on a project with colleagues, enabled the exchange of data with other zooarchaeologists, and it fulfills the need to sustainably store data on the cultural heritage claimed by the UNESCO\footnote{\url{http://www.unesco.org}}. Later, a tunnel also enabled connections from other places as well.

But zooarchaeological data is often gathered in field work, i.e. at remote sites that do not offer a convenient environment for IT services. Thereby, it is typically not possible to enter the data into databases that must be accessed via an internet connection. A synchronization process was required, implementing a Server--Client architecture to ensure working offline at remote places, but also storing data globally, where it can be shared with other users. \cite{lohrer2014}

\begin{figure}[h]
\centering
\vspace{-5pt}\includegraphics[width=1\columnwidth]{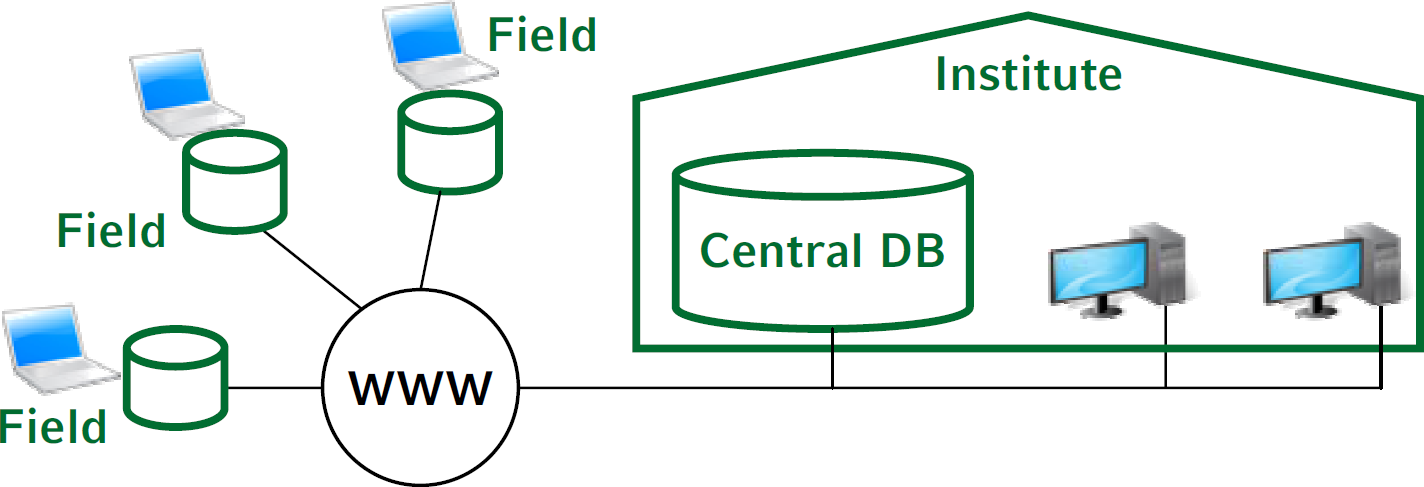}\vspace{-4pt}
\caption{The origin draft for the architecture for the \ossobook synchronization. \cite{kriegel2009} \label{fig:sync_draft1}}
\end{figure}

The first concept for the synchronization was drafted in 2008. ``A client-server architecture ensures that each client [\dots] manages its own local database that is schema-equivalent to the central database at the server. This way, each client can make its updates locally, independently, and -- most important -- offline. At a given time, e.g. when a network connection to the server is established, the client and the server synchronize, i.e. the updates of the client are inserted into the central database at the server and the update of the server'' \cite{kriegel2009}. However, the direct connection to the database without any synchronization would still be possible within the institutes. This architecture is drafted in Fig.~\ref{fig:sync_draft1}. The concept of the synchronization was initially implemented by Jana Lamprecht in 2008 \cite{lamprecht2008}.

%%%%%%%
%%%%%%%

Since this implementation the concept of the synchronization was updated in many development cycles. The synchronization process itself was enhanced and was extended with new features.
% line break in Diss
First of all, the clients do not directly connect to the database anymore, but to a server application that handles the requests and data manipulations. The server application also enables the feature to let the users manage their basic profile information, especially to change and recover their password if needed -- a feature that was lacking before.

Additionally, the possibility to work on different computers with one account was added for a single user. In the origin implementation, this was theoretically also possible, but the technical implementation could cause the loss of data sets during the synchronization process.

The server application was originally written in C/C++, later it was replaced by a PHP program to avoid problems with restrictive firewall settings, that made a communication impossible through other ports than port 80 (HTTP) or 443 (HTTPS). Due to security reasons, all the communication must go through the server application. Therefore, a direct connection to the central database is not supported anymore, not even inside the institutes.

At the same time the logic of the synchronization was also improved and simplified. Also the synchronization panel was extended, it now provides additional information about each project, like the number of entries, the project owner, and how many entries are not synchronized yet.

\begin{figure}[t]
\centering
\includegraphics[width=\columnwidth]{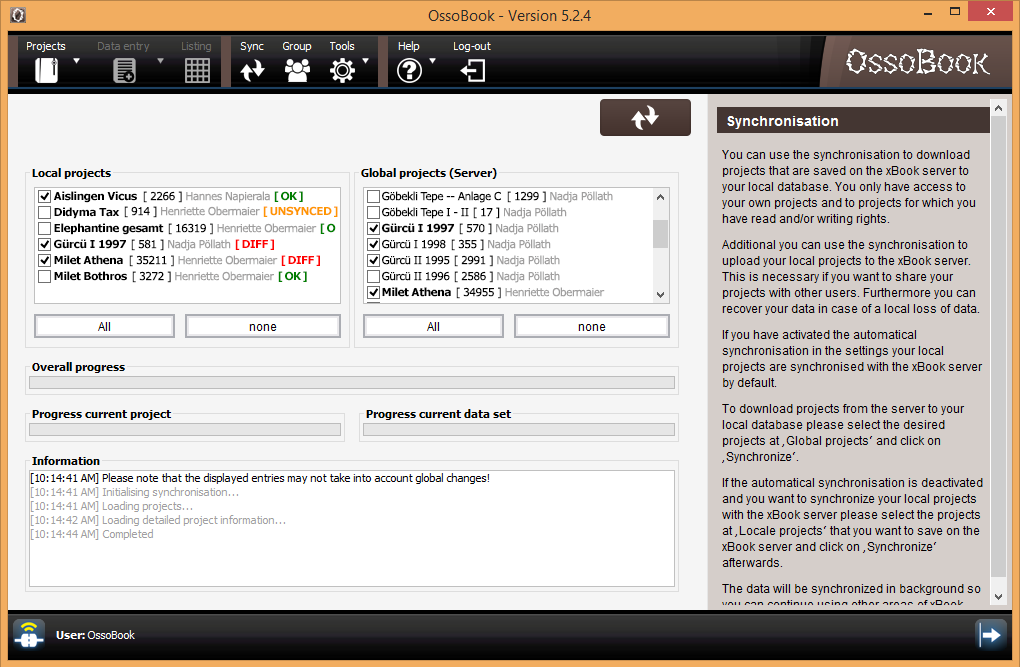}\vspace{-5pt}
\caption{The synchronization panel in \ossobook 5.2.4. \cite{lohrer2016} \label{fig:syncpanel_5-2-4}}
\end{figure}

The detailed implementation of the synchronization is described in the work of Lohrer and Kaltenthaler \cite{lohrer2014}. A screenshot of the latest panel of the synchronization in \ossobook can be viewed in Fig.~\ref{fig:syncpanel_5-2-4}.

%%%%%%%
%%%%%%%

\subsection{Graphical User Interface}

To enable a flexible graphical user interface for the framework and its applications, it was necessary to modify the existing graphical user interface. All static elements had to be replaced with dynamic elements, that can be individually adjusted for each Book. However, elements that are required for each Book have to be integrated to the graphical user interface by default.

In the first step, we replaced the navigation with a new one. The old navigation bar was not designed for elements being dynamically added or updated. It was composed of different images for the normal, hovered, and selected states. The displayed texts on these elements were part of the images, which made translations difficult. For the \xbook framework, the navigation elements are now arranged side by side, are clarified with an individual icon and a short text label. The elements, that were displayed in the sub navigation bar before, can now be accessed by opening a popup menu. All required navigation elements, that are necessary to be accessed in every Book (like `Projects', `Data Entry', `Listing', `Tools', `Help', and `Log-out') are displayed by default. Individual main navigation elements can individually be added by each Book. Adding new elements to the popup menu is supported.

%%%%%%%
\begin{figure*}[t!]
\centering
\includegraphics[width=1\columnwidth]{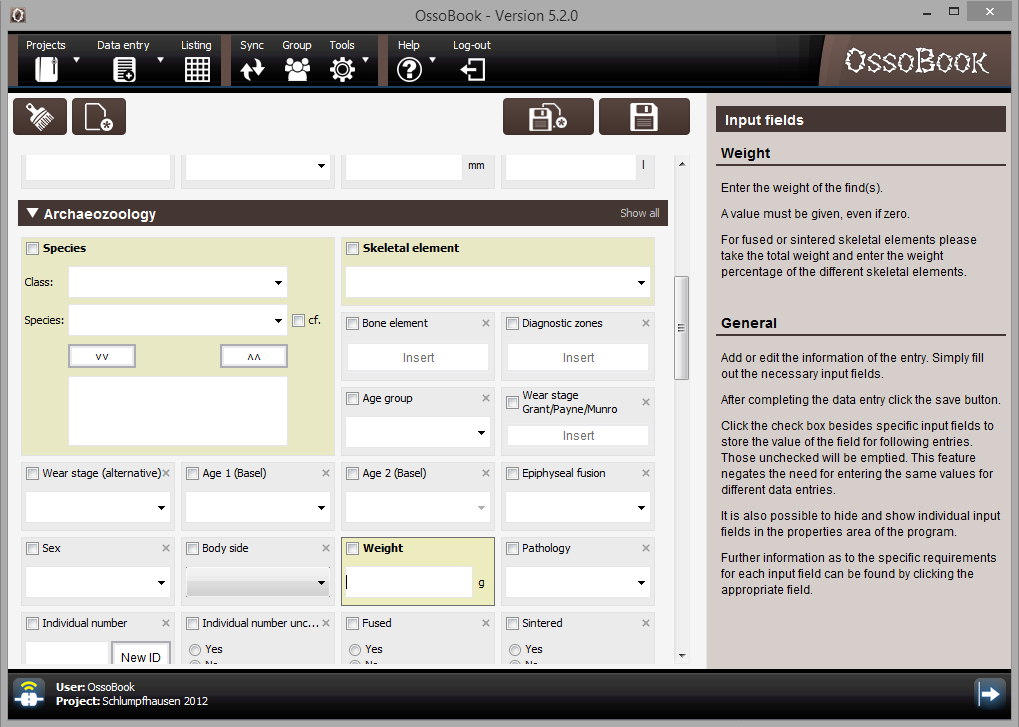}\vspace{-3pt}
\hspace{\fill}
\includegraphics[width=1\columnwidth]{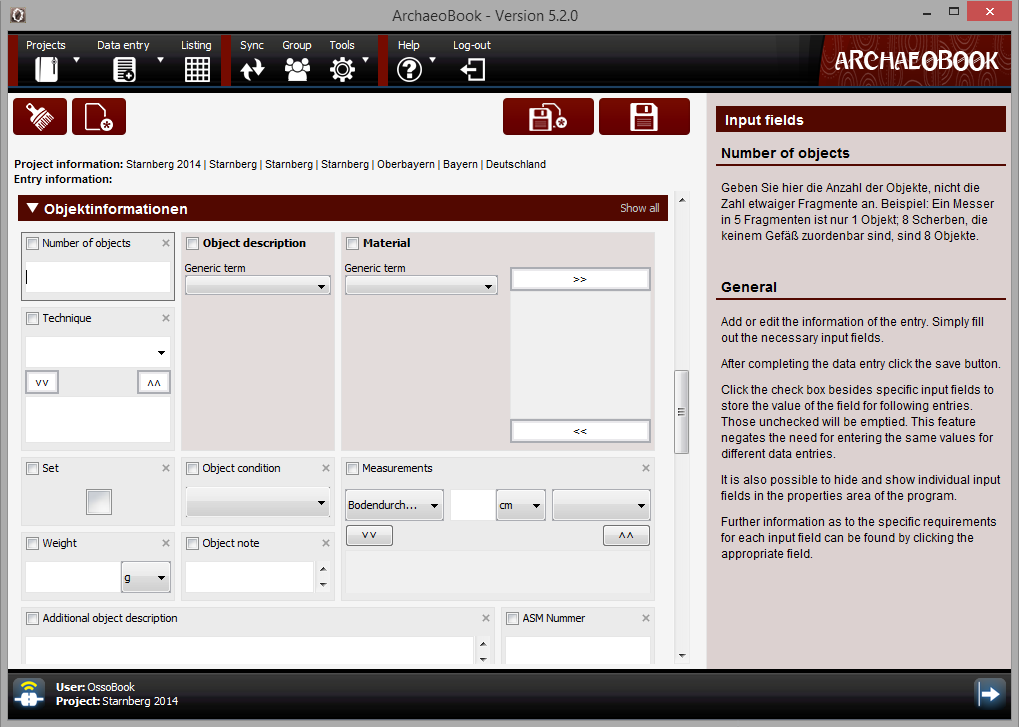}\vspace{-3pt}
\caption{The input mask of \ossobook (top) and \archaeobook (bottom). Both applications are based on the \xbook framework, that provides a basic graphical user interface and functions, but allows customization like e.g. individual input fields. \cite{lohrer2016} \label{fig:xbook_5.2_inputmask}}
\end{figure*}
%%%%%%%

The functionality of the elements of the graphical user interface is fully implemented, but the abstract classes provides methods that have to overridden and implemented by each Book. This way the elements can be customized by defining their contents and configurations. As an example, the input fields -- either the predefined or custom, new ones -- can be defined with their settings and information for the database individually for each Book. The overridden classes define the individual input fields, the logic for handling and displaying the data is defined in the implementation of the \xbook framework. So each Book can have different input fields in the input mask, like indicated in Fig.~\ref{fig:xbook_5.2_inputmask}.

The individualization of other elements of \xbook is similar. The overridden, abstract methods are used to define the information displayed in the project overview screen, the location where to save settings, individual output for the export, the general style of the Book (like the logo, the name of the Book, used colors), and also the definition of the elements displayed in the main and sub navigation.

\subsection{Multiple and crossed-linked input masks}

Before, each Book only had one single input mask with input fields where the users could enter data. During the development of the \xbook framework and its Books it became clear that different, separated input masks are required, especially to avoid redundancy and inconsistency.

So we added the possibility that each Book may have an arbitrary number of input masks with individual input fields. These input masks may be independent from each other, but may also be crossed-linked among each other.

As an example, many archaeological findings are often grouped in boxes or bags, but the information about these boxes and bags should not be entered for each finding again. An own input mask for the boxes or bags is benefiting to avoid extra effort when entering data. Also it is possible that the content of a box or bag changes, so it is easier to select another box or bag instead of re-entering the data for each concerned finding.

The realization of these crossed-linked input masks does not need a complicated architecture of the tables in the database, but requires some more complex SQL queries and adjustments in the graphical user interface, especially for the display and selection in the input mask, the representation in the listing and the export, and adjustments in the synchronization process.

\subsection{Listing and Export}

Besides the data recording, the listing of entered data is also important. The entered data of all input fields must be meaningfully displayed in the data listing, that means the values should be human-readable.

For most of the input fields the value can be directly displayed, like simple text or numeric values. Other, more complex input fields had to be adjusted for the readability, for example by displaying the corresponding text value instead of the corresponding ID, or by displaying a readable date or time format instead of a timestamp. Input fields with multi-inputs should also display all entered values, not only the IDs. For crossed-linked values of other input masks representative text for the crossed-linked entry was necessary as well.

The export of data is also an essential feature. Most of the archaeologists and bioarchaeologists still do their analyses in extern tools or applications like Excel, or want to print them for non-digital archiving. So \xbook provides an export method that saves the data in a file on the local system. There are supported two files systems to export the data: Comma-separated values (csv) and spreadsheet (xls/xlsx) files.

Both the listing and the export have full support for multiple input masks. In the listing a selection of the input mask is available where the user can select which data to be displayed. In the export it is possible so select single input masks to be exported. If a Book with multiple input masks is exported, the data of each mask will be saved in an own csv-file, or in an own sheet of the xls/xlsx file.

\section{Applications using the \xbook framework} \label{applications}

In archaeo-related disciplines, there are many areas with the same or a similar workflow. \xbook offers a framework in which all databases benefit from the provided features. Currently, there are seven different incarnations of the \xbook:

\begin{itemize}

\item \textbf{\ossobook}, a database for zooarchaeological findings, used by
\emph{Bavarian State Collection for Anthropology and Palaeoanatomy Munich}, section Palaeoanatomy\textsuperscript{7},
\emph{ArchaeoBioCenter}\textsuperscript{8}, 
\emph{Institute of Palaeoanatomy, Domestication Research and History of Veterinary Medicine}\textsuperscript{9}, 
\emph{IPNA Basel}\textsuperscript{10},
and members of \emph{BioArch}\textsuperscript{11} and \emph{Deutscher ArchaeoZoologenVerband}\textsuperscript{12}:
    
\item \textbf{\archaeobook}, a database for archaeological findings, used by \emph{Bavarian State Archaeological Collection Munich}\textsuperscript{13}.

\item \textbf{\anthrobook}, a database for anthropological findings, used by \emph{Bavarian State Collection for Anthropology and Palaeoanatomy}, section Anthropology\textsuperscript{7}.

\item \textbf{\excabook}, a database for archaeological findings, used by \emph{Bavarian State Department of Monuments and Sites}\textsuperscript{14}.

\item \textbf{\palaeodepot}, a database for zooarchaeological findings, used by \emph{Bavarian State Collection for Anthropology and Palaeoanatomy Munich}, section Palaeoanatomy\textsuperscript{7}.

\item \textbf{\anthrodepot}, a database for anthropological findings, used by \emph{Bavarian State Collection for Anthropology and Palaeoanatomy Munich}, section Anthropology\textsuperscript{7}.

\item \textbf{\inbook}, a database for archaeological findings, used by \emph{Bavarian State Archaeological Collection Munich}\textsuperscript{13}.

\end{itemize}

\vspace{0.2cm}

\begin{footnotesize}
\noindent \textsuperscript{7}\url{http://www.snsb.mwn.de}\\
\textsuperscript{8}\url{http://www.archaeobiocenter.uni-muenchen.de}\\
\textsuperscript{9}\url{http://www.palaeo.vetmed.uni-muenchen.de}\\
\textsuperscript{10}\url{http://ipna.unibas.ch}\\
\textsuperscript{11}\url{http://www.archeorient.mom.fr/recherche-et-activites/participation-a-des-reseaux/GDRE-BIOARCH}\\
\textsuperscript{12}\url{http://www.archaeozoologenverband.de}\\
\textsuperscript{13}\url{http://www.archaeologie-bayern.de}\\
\textsuperscript{14}\url{http://www.blfd.bayern.de}
\end{footnotesize}
\section{Availability} \label{availability}

The \xbook framework is available from the URL:\\
\url{http://xbook.vetmed.uni-muenchen.de/download/}

% \fi
\IEEEpeerreviewmaketitle

%\newpage



\begin{thebibliography}{1}
\bibitem {ossobook}
Kaltenthaler,~D.; Lohrer,~J.-Y.; Kr\"oger,~P.; van der Meijden,~C.; Granado,~E.; Lamprecht,~J.; N\"ucke,~F.; Obermaier,~H.; Stopp,~B.; Baly,~I.; Callou,~C.; Gouriichon,~L.; Peters,~J.; P\"ollath,~N.; Schibler,~J.: OssoBook v5.6.0. [Computer Software] Munich, Germany; Basel, Switzerland (2018), \url{http://xbook.vetmed.uni-muenchen.de}.

\bibitem {archaeobook}
Kaltenthaler,~D.; Lohrer,~J.-Y.; Kr\"oger,~P.; van der Meijden,~C.; Harrington,~C.; Claßen,~E.; Gebhard,~R.; Marzinzik,~S.; Schwarzberg,~H.: ArchaeoBook v5.6.0. [Computer Software] Munich, Germany (2018), \url{http://xbook.vetmed.uni-muenchen.de}.

\bibitem {anthrobook}
Kaltenthaler,~D.; Lohrer,~J.-Y.; Kr\"oger,~P.; van der Meijden,~C.; Sizova,~T.; M\"osch,~A.; Harbeck,~M.; Grigat,~A.; Toncala,~A.: ArchaeoBook v5.6.0. [Computer Software] Munich, Germany (2018), \url{http://xbook.vetmed.uni-muenchen.de}.

\bibitem {excabook}
Kaltenthaler,~D.; Lohrer,~J.-Y.; Kr\"oger,~P.; van der Meijden,~C.; Wanke,~T.; Sassen,~I.; Jantos,~S.; Rahm,~A.; Wanninger,~R.; Haberstroh,~J.; Sommer,~S.: ExcaBook v5.6.0. [Computer Software] Munich, Germany (2018), \url{http://xbook.vetmed.uni-muenchen.de}.

\bibitem {palaeodepot}
Kaltenthaler,~D.; Lohrer,~J.-Y.; Kr\"oger,~P.; van der Meijden,~C.; Obermaier,~H.: PalaeoDepot v5.6.0. [Computer Software] Munich, Germany (2018), \url{http://xbook.vetmed.uni-muenchen.de}.

\bibitem {anthrodepot}
Kaltenthaler,~D.; Lohrer,~J.-Y.; Kr\"oger,~P.; van der Meijden,~C.; Harbeck,~M.; Grigat,~A.: AnthroDepot v5.6.0. [Computer Software] Munich, Germany (2018), \url{http://xbook.vetmed.uni-muenchen.de}.

\bibitem {inbook}
Kaltenthaler,~D.; Lohrer,~J.-Y.; Kr\"oger,~P.; van der Meijden,~C.; Harrington,~C.; Claßen,~E.; Gebhard,~R.; Marzinzik,~S.; Schwarzberg,~H.: InBook v5.6.0. [Computer Software] Munich, Germany (2018), \url{http://xbook.vetmed.uni-muenchen.de}.

\bibitem {kaltenthaler2012}
Kaltenthaler,~D.: Visual Cluster Analysis of the Archaeological Database {OssoBook} with special focus on Data Integrity and Consistency. Diploma thesis, Ludwig-Maximilians-Universit{\"{a}}t M{\"{u}}nchen, Germany, 2012.

\bibitem {lohrer2012}
Lohrer,~J.-Y.: Density Based Cluster Analysis of the Archaeological Database {OssoBook} in Condideration of Aspects of Data Quality. Diploma thesis, Ludwig-Maximilians-Universit{\"{a}}t M{\"{u}}nchen, Germany, 2012.

\bibitem {schibler1998}
Schibler,~J.: {OSSOBOOK}, a database system for archaeozoology. In: Anreiter,~P.; Bartosiewicz,~L.; Jerem,~E.; Meid,~W.: Man and the Animal World: Studies in Archaeozoology, Archaeology, Anthropology and Palaeolinguistics in Memoriam S\'andor B\"ok\"onyi. Archaeolingua, No. 8, pp. 491--510, 1998.

\bibitem {lamprecht2008}
Lamprecht,~J.: Conception and Implementation of a Intermittently Synchronized Database System for Palaeoanatomic applications. Diploma thesis, Ludwig-Maximilians-Universit{\"{a}}t M{\"{u}}nchen, Germany, 2008.

\bibitem {kaltenthaler2011}
Kaltenthaler,~D.: Design and Implementation of a Graphical User Interface for the Archaeozoological Database {OssoBook}. Project thesis, Ludwig-Maximilians-Universit{\"{a}}t M{\"{u}}nchen, Germany, 2011.

\bibitem {lohrer2011}
Lohrer,~J.-Y.: Design and Implementation of a Dynamic Database for Archaeozoological Applications. Project thesis, Ludwig-Maximilians-Universit{\"{a}}t M{\"{u}}nchen, Germany, 2011.

\bibitem {danti2010}
Danti,~S.: Cluster Analysis of Features of Animal Bones and Similarity Search on Multi Instance Objects of the Archaeozoological Data Pool. Diploma thesis, Ludwig-Maximilians-Universit{\"{a}}t M{\"{u}}nchen, Germany, 2010.

\bibitem {tsukanava2010}
Tsukanava,~Y.: Development and Appliance of Data Mining Methods on the Palaeoanatomic Data Collection. Diploma thesis, Ludwig-Maximilians-Universit{\"{a}}t M{\"{u}}nchen, Germany, 2010.

\bibitem {neumayer2012}
Neumayer,~T.: Design and Implementation of Analysis Methods for Archaeozoological Data. Bachelor thesis, Ludwig-Maximilians-Universit{\"{a}}t M{\"{u}}nchen, Germany, 2012.

\bibitem {oracle2017}
Oracle: The Java Tutorials: Serializable Objects. [Online] Accessed at \url{https://docs.oracle.com/javase/tutorial/jndi/objects/serial.html} on December 06, 2017.

\bibitem {sockets2018}
The GNU C Library: Sockets. [Online] Accessed at \url{http://www.gnu.org/savannah-checkouts/gnu/libc/manual/html_node/Sockets.html} on January 23, 2018.

\bibitem {garg2002}
Garg,~R.~P.; Sharapov,~I.: Techniques for Optimizing Applications: High Performance Computing. Prentice Hall Professional Technical Reference, 2002.

\bibitem {goetz2002}
Goetz,~B.: Java theory and practice: Thread pools and work queues. [Online] Accessed at \url{https://www.ibm.com/developerworks/java/library/j-jtp0730/index.html} on December 06, 2017.

\bibitem {goll2014}
Goll,~J.: Architektur- und Entwurfsmuster der Softwaretechnik. Springer, 2014.

\bibitem {json1}
The JSON Data Interchange Syntax: Standard ECMA-404, 2nd Edition / December 2017. \url{http://www.ecma-international.org/publications/standards/Ecma-404.htm}

\bibitem {json2}
RFC 8259: The JavaScript Object Notation (JSON) Data Interchange Format. [Online] Accessed at \url{https://tools.ietf.org/html/rfc8259} on January 23, 2018.

\bibitem {jsoncompression}
McAnlis,~C.: JSON Compression: Transpose \& Binary. [Online] Accessed at \url{http://mainroach.blogspot.de/2013/08/json-compression-transpose-binary.html} on January 23, 2018.

\bibitem {xfjson}
JFJSON. [Online] Accessed at \url{https://github.com/mainroach/compression/tree/master/xfjson} on January 23, 2018.

\bibitem {flatbuffers}
Google Developers: FlatBuffers. [Online] Accessed at \url{https://google.github.io/flatbuffers/} on January 23, 2018.

\bibitem {protocolbuffers}
Google Developers: Protocol Buffers. [Online] Accessed at \url{https://developers.google.com/protocol-buffers/} on January 23, 2018.

\bibitem {bson}
BSON. [Online] Accessed at \url{http://bsonspec.org/} on January 23, 2018.

\bibitem {lohrer2014}
Lohrer,~J.-Y.; Kaltenthaler,~D.; Kr\"oger,~P.; van der Meijden,~C.; Obermaier,~H.: A Generic Framework for Synchronized Distributed Data Management in Archaeological Related Disciplines. In: 10th {IEEE} International Conference on e-Science, eScience 2014, S\~ao Paulo, Brazil, October 20-24, 2014, pp. 5--12. Springer, 2014.

\bibitem {kriegel2009}
Kriegel,~H.-P.; Kr\"oger,~P.; Obermaier,~H.; Peters,~J.; Renz,~M.; van der Meijden,~C.: {OSSOBOOK:} database and knowledgemanagement techniques for archaeozoology. In: Proceedings of the 18th {ACM} Conference on Information and Knowledge Management, {CIKM} 2009, Hong Kong, China, November 2-6, 2009, pp. 2091--2092. 2014.

\bibitem {lohrer2016}
Lohrer,~J.-Y.; Kaltenthaler,~D.; Kr\"oger,~P.; van der Meijden,~C.; Obermaier,~H.: A Generic Framework for Synchronized Distributed Data Management in Archaeological Related Disciplines. In: Future Generation Comp. Syst., Vol. 56, pp. 558--570. 2016.
\end{thebibliography}
\end{document}